\documentclass[twocolumn]{aastex6}

\usepackage{graphicx}
\usepackage{amsmath}
\usepackage{amssymb}
\usepackage{hyperref}
\usepackage{nicefrac}
\usepackage{xcolor}
\usepackage{cancel}

\begin{document}

\title{Equation of state constraints from nuclear physics, neutron star masses, and\\
future moment of inertia measurements}

\author{S.\ K.\ Greif,$^{1,2}$ K.\ Hebeler,$^{1,2}$ J.\ M.\ Lattimer,$^3$ C.\ J.\ Pethick,$^{4,5}$ 
and A.\ Schwenk$\,^{1,2,6}$}
\affil{
$^1$Institut f\"ur Kernphysik, Technische Universit\"at Darmstadt, D-64289 Darmstadt, Germany \\
$^2$ExtreMe Matter Institute EMMI, GSI Helmholtzzentrum f\"ur Schwerionenforschung GmbH, D-64291 Darmstadt, Germany \\
$^3$Department of Physics and Astronomy, Stony Brook University, Stony Brook, New York 11794-3800, USA \\
$^4$The Niels Bohr International Academy, The Niels Bohr Institute, \\
University of Copenhagen, Blegdamsvej 17, DK-2100 Copenhagen \O, Denmark \\
$^5$NORDITA, KTH Royal Institute of Technology and Stockholm University,
Roslagstullsbacken 23, SE-10691 Stockholm, Sweden \\
$^6$Max-Planck-Institut f\"ur Kernphysik, Saupfercheckweg 1, D-69117 Heidelberg, Germany
}

\begin{abstract}
We explore constraints on the equation of state (EOS) of neutron-rich matter based
on microscopic calculations up to nuclear densities and observations
of neutron stars. In a previous work we showed that
predictions based on modern nuclear interactions derived within chiral
effective field theory and the observation of two-solar-mass neutron stars 
result in a robust uncertainty range for neutron
star radii and the EOS over a wide range of densities. In
this work we extend this study, employing both the piecewise polytrope
extension from Hebeler et al. as well as the speed of sound model 
of Greif et al., and show that moment of inertia
measurements of neutron stars can significantly improve the constraints
on the EOS and neutron star radii.
\end{abstract}


\maketitle

\section{Introduction}
\label{sect:Intro}

Recently, there has been significant progress in our understanding
of the equation of state (EOS) of dense matter. This was triggered by
advances in nuclear theory, new constraints from precise measurements
of heavy neutron stars, as well as astrophysical observations from the
LIGO/Virgo~\citep{LIGO18NSradii,LIGO19update} and 
NICER~\citep{Mill19NICER,Raai19NICEREOS,Rile19NICER} collaborations.
These offer complementary insights to the EOS. While
nuclear theory provides reliable predictions for neutron-rich matter up to densities 
around saturation density (\mbox{$\rho_0 = 2.8 \times 10^{14} \,
\text{g} \, \text{cm}^{-3}$}), observations of neutron stars, and neutron
star mergers probe the EOS over a higher range of densities but provide
indirect constraints.

In nuclear physics the development of chiral effective field theory (EFT) has
revolutionized our approach to nuclear forces. The description of the
interactions between neutrons and protons, both particles with a complex
substructure, has been a challenge in nuclear theory for decades. Pioneered by
the seminal works of \citet{Wein90NFch,Wein91chNp}, chiral EFT has
now become the only known framework that allows a systematic expansion of
nuclear forces at low energies~\citep{Epel09RMP,Mach11PR,Hamm13RMP} based on the
symmetries of quantum chromodynamics (QCD), the fundamental theory of the strong
interaction. In addition, chiral EFT allows one to derive systematic estimates of
uncertainties for observables. Incorporating such chiral EFT interactions in
microscopic many-body frameworks makes it possible to compute uncertainty bands for the
pressure and energy density of matter~\citep{Hebe10nmatt,Carb13nm,
Holt13PPNP,Tews13N3LO,Well14nmtherm,
Dris16asym,Dris19MCshort,Dris20nmBayes,Lynn16QMC3N}. As any effective
low-energy theory, chiral EFT contains an intrinsic breakdown scale. When
approaching this breakdown scale with increasing energy or density the
convergence of the effective expansion becomes slower until eventually it
breaks down. This breakdown scale translates into an upper density limit for
such calculations. The precise value for this upper density limit is
still unknown, and also depends on details of the interactions. In a previous
work~\citep{Hebe13ApJ}, we chose an upper density limit of $1.1 \, \rho_0$ for
neutron-rich matter. This limit represents a rather conservative choice and it
might be possible to push this limit to somewhat higher densities~\citep{Tews18cs},
although a full understanding of the implied uncertainties is still
an open problem. Finally, for very high densities ($\rho \gtrsim 50 \, \rho_0$),
there are model-independent constraints from perturbative QCD 
calculations of quark matter~\citep{Kurk10pQCD}.

Neutron star observations provide powerful constraints on the EOS
beyond the densities accessible by nuclear theory as well as
laboratory experiments~\citep{Tsan12esymm}.  In particular, the
precise mass measurements of the pulsars PSR J1614-2230 and PSR
J0348+0432 with masses of $1.928 \pm 0.017 \,
M_\odot$~\citep{Fons16pulsars} and 
$2.01 \pm 0.04 \, M_\odot$~\citep{Anto13PSRM201}
turned out to be
a key discovery, as the existence of such heavy neutron stars puts
tight constraints on the EOS and the composition of matter, ruling out
a large number of EOSs with simple inclusion of exotic degrees of
freedom like hyperons or deconfined quarks. Recently, 
the mass of the pulsar PSR J0740+6620 was measured to be
$2.14 \substack{+0.10 \\ -0.09} \, M_\odot$~\citep{Crom19massive},
which further tightens these constraints.

In this work, we study the EOS constraints that can be achieved from future moment 
of inertia measurements, in addition to the heavy mass
constraint discussed above. The moment of inertia has been suggested
to provide  complementary constraints for the 
EOS~\citep{Rave94moi,Lyne04ns,Latt05nsMOI}. It
can be obtained from measurements of the rate of advance of the
periastron, $\dot{\omega}$~\citep{Damo88pulsar}. This advance is mainly caused
by the relativistic spin-orbit coupling in a binary 
system~\citep{Bark75socoup,Wex95binary,Kram09DNS}, and the magnitude of the
advance depends sensitively on the orbital period and the compactness of the
binary system. In 2003, the double neutron-star system PSR J0737--3039 was
discovered~\citep{Burg03ns,Lyne04ns}. This system is particularly promising
for such measurements, as it is extremely compact with an orbital period of only
2.4~hr~\citep{Burg03ns,Burg05DNS,Lyne04ns}. In addition, due to the high orbital
inclination~\citep{Burg03ns,Burg05DNS}, the masses of the two neutron stars
have been determined very precisely to be \mbox{$1.3381(7) \, M_\odot$} and
\mbox{$1.2489(7) \, M_\odot$}~\citep{Kram09DNS}. Due to the compactness of the
system, the moment-of-inertia correction to $\dot{\omega}$ is estimated to be
an order of magnitude larger for PSR J0737-3039A (the heavier of the two pulsars)
than for other systems like PSR B1913+16~\citep{Lyne04ns}. Such a moment of
inertia measurement has to be performed over a
long period of time and an increase of timing precision would be
beneficial~\citep{Kram09DNS}. Based on this, it was argued that a moment of inertia
measurement with a relative uncertainty of about \mbox{$10\%$} may be
achievable~\citep{Damo88pulsar,Latt05nsMOI,Kram09DNS}.

Previous works studied to what extent such measurements are able to provide
constraints for different types of EOS~\citep{Morr04nsMOI,Bejg05nsMOI,%
Latt05nsMOI}. In particular, \citet{Rave94moi} showed that the moment of inertia
can be parameterized efficiently as a function of the compactness parameter, and
\citet{Latt05nsMOI} demonstrated that a universal relation between the moment of inertia
and the compactness parameter exists, which can be used to provide constraints
on neutron star radii. More recently, \citet{Stei15MOI}, \cite{Gord16MOIQCD},
and \citet{Lim18MOI} studied the moment of inertia based on
neutron star observations and EOS constraints, and \citet{Rait16nsMOI}
investigated the inference of neutron star radii from moment of inertia measurements.

In this work, we study how microscopic calculations based on chiral EFT
interactions combined with neutron star masses and a future moment of inertia
measurement can provide novel predictions for the EOS and neutron star 
radii. In Section~\ref{sect:MOI_EFT}, we briefly review our approach employing
both the piecewise polytrope extension from~\citet{Hebe13ApJ} as well as the
speed of sound model of~\citet{Greif19MNRAS} and present uncertainty ranges for neutron star observables such as the mass, the radius, and the moment of inertia. In Section~\ref{sect:constraints}, we present
our results for neutron star radii, and how these can improve upon information
from the neutron star merger GW170817~\citep{LIGO19update}. Moreover, we discuss the resulting EOS constraints and explore scaling relations for the dimensionless moment of inertia. Finally, we conclude in
Section~\ref{sect:summary}.

\section{Constraints from nuclear theory and neutron star masses}
\label{sect:MOI_EFT}

In \citet{Hebe10PRL,Hebe13ApJ} we combined constraints from nuclear
physics and neutron star masses to derive constraints for the
EOS for all densities relevant for neutron stars. We
briefly review the strategy of this work and refer
to \citet{Hebe13ApJ} for details:

a) The first constraint results from microscopic calculations of neutron-rich
matter up to density $\rho_1 = 1.1 \, \rho_0$ based on modern nuclear interactions
derived from chiral EFT~\citep{Hebe10nmatt,Tews13N3LO}. These calculations
resulted in uncertainty bands for the energy density and pressure. For densities
below $\rho_{\text{crust}} = 0.5 \, \rho_0$ the BPS crust EOS of
\citet{Baym71BPS} and \citet{Nege73BPS} was used. Remarkably, around the
transition density $\rho_{\text{crust}}$ both EOSs overlap smoothly,
so that our final results are insensitive to the particular choice
for $\rho_{\text{crust}}$.

\begin{figure*}[t]
\centering
\includegraphics[width=\textwidth]{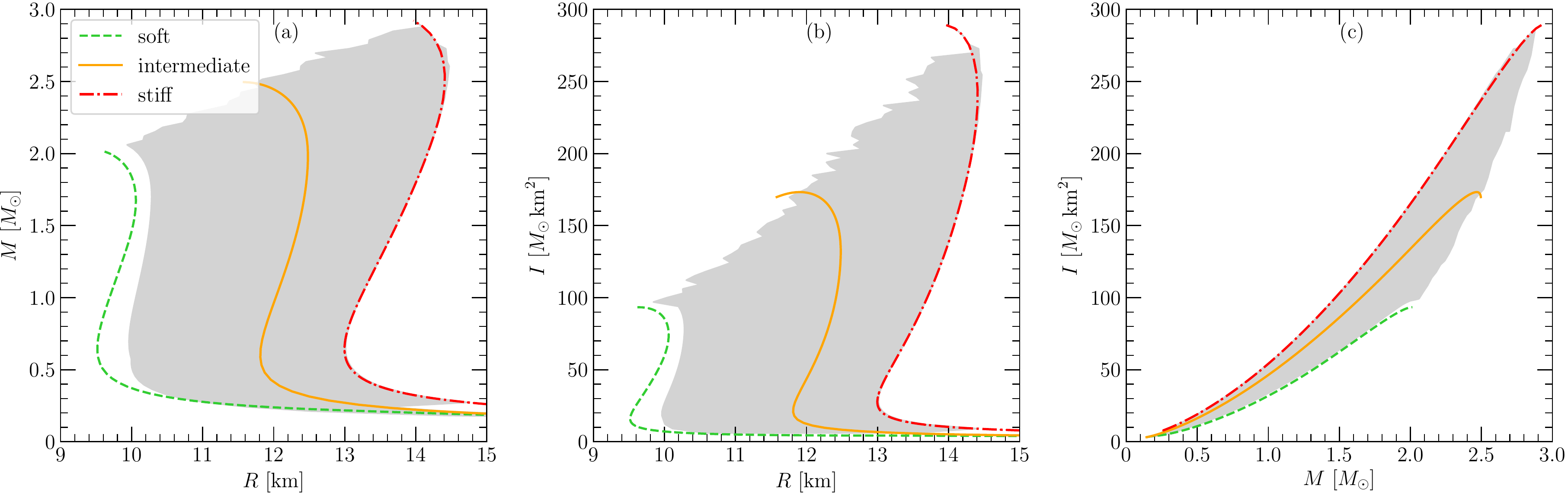}
\caption{Results for mass $M$, radius $R$, and moment of inertia $I$ of neutron
stars based on the EOS constraints (bands) derived with the piecewise
polytrope model based on chiral EFT calculations up to density $\rho_1 = 1.1 \, \rho_0$, the new mass
constraint $M_{\text{obs}} \geqslant 2.05 \, M_{\odot}$, and causality constraints.
The individual panels (a), (b), and (c) show the mass-radius, moment of
inertia-radius, and moment of inertia-mass results, respectively. The
green (dashed), yellow (solid), and red (dotted-dashed) lines correspond to the three
representative EOS (soft, intermediate, and stiff, respectively) from 
\citet{Hebe13ApJ}. Note that the latter are for the old
mass constraint $M_{\text{obs}} \geqslant 1.97 \, M_{\odot}$, so that the
soft EOS leads to smaller radii.}
\label{fig:MRI}
\end{figure*}

b) Based on the constraints from nuclear physics at low densities the EOS
was extended in a general way to higher densities using piecewise polytropes,
\mbox{$P(\rho)=K_i \rho^{\Gamma_i}$}, with the adiabatic indices $\Gamma_i$ and
constants $K_i$ (see also \citet{Read09polyEOS}). The values for $\Gamma_i$
are allowed to vary freely, whereas the values of $K_i$ are fixed by the constraint
that the EOS should be continuous as a function of density. For the extension
beyond $\rho_1$, three polytropes characterized by exponents $\Gamma_1$,
$\Gamma_2$ (beyond $\rho_{12}$), and $\Gamma_3$ (beyond $\rho_{23}$) allow
one to control the softness or stiffness of the EOS in a given density region, and
the transition densities $\rho_{12}$ and $\rho_{23}$ between polytropes are
allowed to vary as well. Sampling all possible EOSs using the step size $\Delta 
\Gamma_i = 0.5$ and $\Delta \rho_{12,23} = \rho_0/2$ results in a very large
number of possible EOSs (for details, see \citet{Hebe13ApJ}), which include
constructions that mimic first-order phase transitions. The values of
$\Gamma_i$, $\rho_{12}$, and $\rho_{23}$ are then constrained by the condition that each EOS must be able to support a neutron star of at least 
$M_{\text{obs}} = 2.05 \, M_\odot$, which we take as the 68\% lower limit of the mass of the heaviest precisely
known pulsar~\citep{Crom19massive}. This mass constraint provides an
update compared to the $1\sigma$ lower limit ($1.97 \, M_\odot$) of the 
mass of PSR J0348+0432~\citep{Anto13PSRM201} used in \citet{Hebe13ApJ}.

c) As the final constraint we require that the speed of sound, $c_\text{s}$, remain
smaller than the speed of light, $c$, for all densities: $c_{\text{s}}/c =
\sqrt{dP /d\mathcal{E}} \leqslant 1$, where $P$ is the pressure and $\mathcal{E}$
is the energy density. Each EOS is followed in density until causality is violated
or the maximum neutron star mass is reached when $dM/dR = 0$.

The combination of these three conditions leads to mass-radius constraints on
neutron stars shown in panel~(a) of Fig.~\ref{fig:MRI}. In general, the
boundaries of the band are spanned by a large number of different EOSs,
but to distinguish soft and stiff EOSs, we show the three representative
EOSs (soft, intermediate, and stiff) of \citet{Hebe13ApJ}, which span the
radius range as shown in Fig.~\ref{fig:MRI}, while the soft EOS leads to
somewhat smaller radii due to the previous mass constraint $M_{\text{obs}} 
\geqslant 1.97 \, M_{\odot}$. For a typical \mbox{$M=1.4 \, M_{\odot}$}
star, the update gives a radius range of \mbox{$R= 10.2$--$13.6 \, \text{km}$}
(taking the chiral EFT constraints from renormalization-group-evolved interactions, 
which have improved many-body convergence; \citet{Hebe13ApJ}).

In order to explore the sensitivity to details of the high-density extension,
we also employ the speed of sound model of \citet{Greif19MNRAS} in
addition to the piecewise polytrope extension. The speed of sound model
is based on the same crust EOS and chiral EFT band, but uses a parameterization
of the speed of sound to high densities, which includes a maximum in the speed
of sound $c_s^2/c^2 > 1/3$ and an asymptotic convergence to the 
conformal limit from below, for very high densities ($\rho \gtrsim 50 \, \rho_0$)
suggested by the perturbative QCD calculations~\citep{Kurk10pQCD}. The two different 
extensions lead to small changes in the predicted ranges, e.g., for the radius of a neutron star. 
These differences result from the choice of three polytropes and the particular functional form chosen 
for the speed of sound parameterization, and would be diminished for arbitrarily fine discretizations 
of the high-density part of the EOS.

In this work we build on our past mass-radius results~\citep{Hebe13ApJ,Greif19MNRAS} 
and investigate how future moment of inertia
measurements of neutron stars will be able to further constrain the EOS and neutron
star radii. To this end, we investigate rotating neutron stars and use the Hartle--Thorne
slow-rotation approximation~\citep{Hart67slowrot,Hart68slowrot}. Later studies
have been more conservative, verifying the applicability of this treatment to frequencies up to \mbox{$f \approx 200 \, \text{Hz}$} \citep{Benh05slowrot, Cipo15slowrot}. The heavier neutron star of the system PSR J0737--3039 has a period of about
23~ms~\citep{Burg03ns,Lyne04ns} and can hence reliably be treated within
the slow-rotation approximation.

\begin{figure}[t]
\centering
\includegraphics[width=0.45\textwidth]{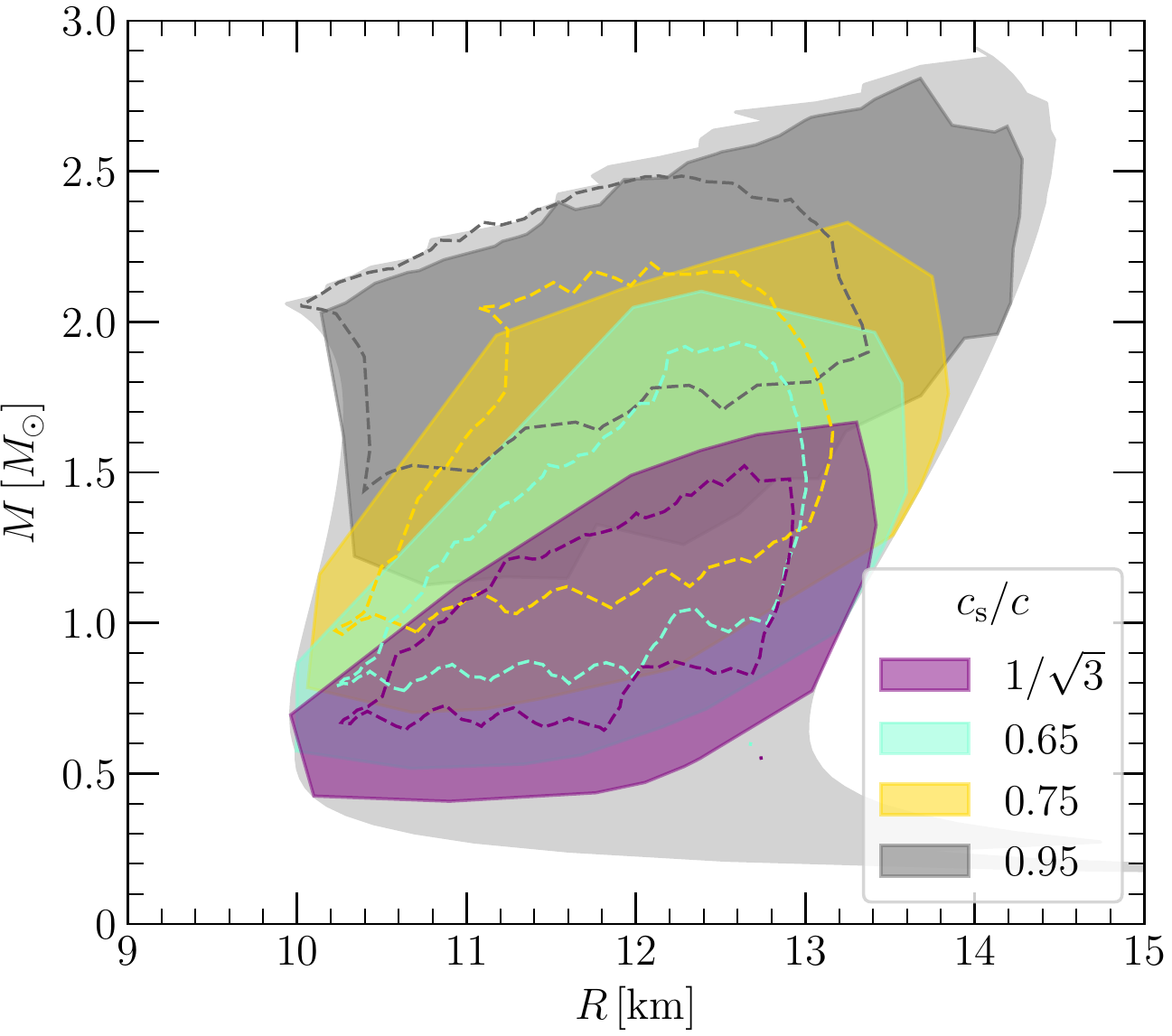}
\caption{Mass $M$ as a function of radius $R$. The gray area depicts the
entire region allowed by the general EOS construction using the piecewise polytrope
extension. The highlighted areas represent $M$--$R$ pairs that reach values for the 
speed of sound $c_{\text{s}}/c \leqslant 1/\sqrt{3}$ (purple), 0.65 (blue), 0.75 (orange), 
and 0.95 (dark gray). The dashed lines mark the corresponding regions for the 
speed of sound model.}
\label{fig:speed_of_sound}
\end{figure}

\begin{figure*}[t]
\centering
\includegraphics[width=\textwidth]{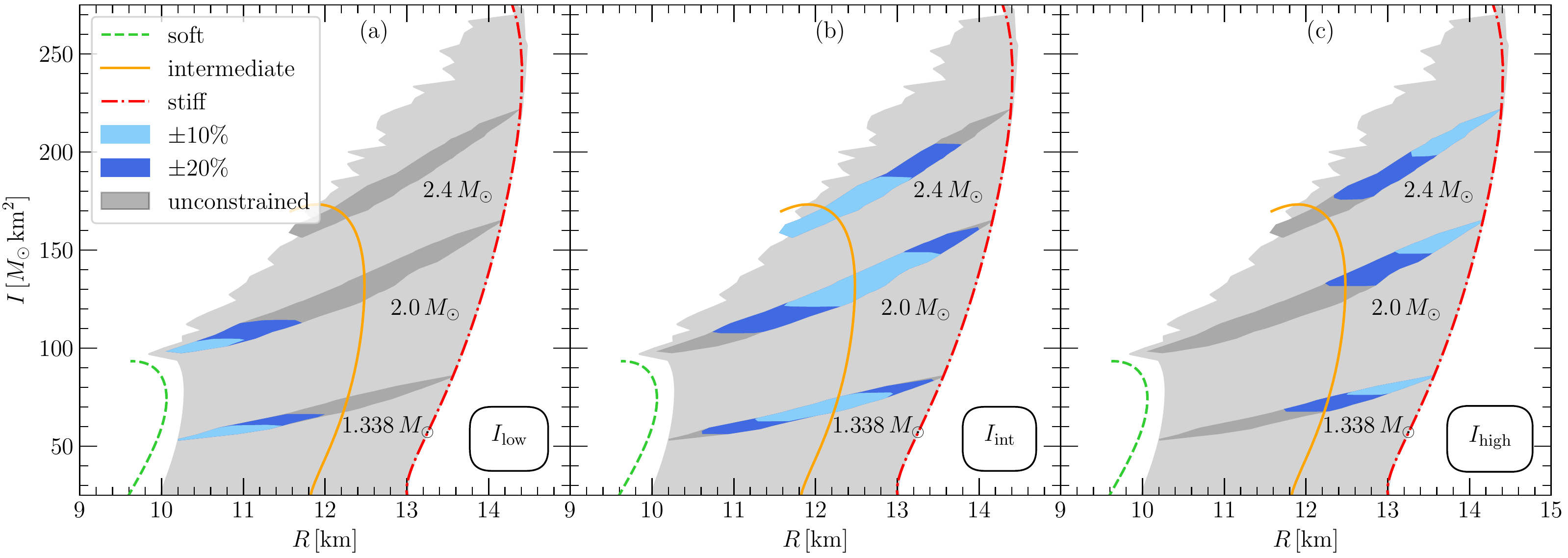}
\caption{Moment of inertia $I$ as a function of radius $R$. The gray band
gives the allowed $I$--$R$ range resulting from the general EOS band
for the piecewise polytrope extension as shown in Fig.~\ref{fig:MRI}. The dark gray, light blue, and
dark blue areas show the allowed $I$--$R$ values for the particular neutron
star masses indicated, where the dark gray area includes all possible $I$--$R$
pairs for each mass, and the light blue (dark blue) area corresponds
to an assumed measurement of the moment of inertia with central value
$I_c$ given in Table~\ref{tab:radius_ranges} with a relative uncertainty of
$\Delta I = \pm 10\%$ $(\pm 20\%)$. The three panels assume central
values $I_c$ that approximately correspond to the soft (a), intermediate (int) (b), 
and stiff (c) EOS, see Table~\ref{tab:radius_ranges}. Note that for a $2.4 \, 
M_\odot$ neutron star, the soft EOS is ruled out
and thus no compatible $I_c$ exists in this case.}
\label{fig:radius_constraints}
\end{figure*}

Panels (b) and (c) of Fig.~\ref{fig:MRI} show the results for the moment of
inertia $I$ as a function of neutron star mass and radius based on our EOS
bands from the piecewise polytrope extension. 
The moment of inertia can reach values up to about $290 \, M_{\odot} 
\, \text{km}^2$ for very heavy neutron stars, where the maximal values 
are clearly correlated with the stiffness of the EOS. In addition, it is manifest
that the three EOSs which are representative with respect to the radius are
also representative with respect to the moment of inertia and practically span
the full moment-of-inertia range (with only minor modifications for the soft
EOS due to the new mass constraint). For the pulsar PSR J0737-3039A with 
$M = 1.338 \, M_\odot$ we find the moment of inertia to be in the range
$I = 53.2$--$85.7 \, M_\odot \, \text{km}^2$. Our predicted range is significantly 
smaller than that of \citet{Rait16nsMOI}, where $I = 21.1$--$113.2 \,
M_\odot \, \text{km}^2$, and similar to the range obtained by
\citet{Gord16MOIQCD} with $I= 60.3$--$90.5 \, M_\odot \, \text{km}^2$.

In addition, we show the speed of sound $c_\text{s}$ reached in
our general EOS bands. In Fig.~\ref{fig:speed_of_sound} the highlighted areas
represent $M$--$R$ pairs that reach particular values for $c_{\text{s}}/c$.
Note that $c_\text{s}/c$ is small at low densities in the nonrelativistic
chiral EFT calculations and reaches $1/\sqrt{3} \approx 0.577$ from below in the perturbative
QCD regime~\citep{Kurk10pQCD}. Figure~\ref{fig:speed_of_sound} clearly demonstrates
that $c_\text{s}/c$ has to reach values of around 0.65 to be compatible with
two-solar-mass neutron stars. In particular, if one demands that $c_\text{s}/c 
\leqslant 1/\sqrt{3}$ for all densities in neutron-star matter, no EOS exists in
our general construction that is compatible with the observed heavy neutron
stars. This has also been pointed out by \citet{Beda15cs} and is consistent
with the findings of \citet{Tews18cs} and \citet{Greif19MNRAS}.

\begin{table}[t]
\begin{center}
\caption{Radius Constraints Resulting from Mass and
Moment of Inertia Measurements for the Same Star, Assuming the Mass Uncertainty
Is Negligible and Using the Piecewise Polytrope Extension}
\label{tab:radius_ranges}
\begin{tabular}{cccccc}
$M$ & & $I_c$ & $R(\pm 10\%)$ & $R(\pm 20\%)$ & $R$ \\ 
\hline \hline
& $I_{\text{low}}$ & 55 & 10.2--11.4 & 10.2--12.0 & 10.2--13.6 \\
1.338 & $I_{\text{int}}$ & 70 & 11.3--12.9 & 10.6--13.4 & 10.2--13.6 \\ 
& $I_{\text{high}}$ & 85 & 12.5--13.6 & 11.8--13.6 & 10.2--13.6 \\ \hline
& $I_{\text{low}}$ & 95 & 10.1--11.0 & 10.1--11.7 & 10.1--14.2 \\ 
2.0 & $I_{\text{int}}$ & 135 & 11.6--13.5 & 10.8--14.0 & 10.1--14.2 \\ 
& $I_{\text{high}}$& 165 & 13.1--14.2 & 12.3--14.2 & 10.1--14.2 \\ \hline
& $I_{\text{low}}$ & -- & -- & -- & 11.6--14.4 \\ 
2.4 & $I_{\text{int}}$ & 170 & 11.6--13.2 & 11.6--13.8 & 11.6--14.4 \\ 
& $I_{\text{high}}$ & 220 & 13.3--14.4 & 12.4--14.4 & 11.6--14.4 \\ \hline \hline
\end{tabular} 
\end{center}
\textbf{Note.} The columns give the assumed values for $M$ (in units
of $M_{\odot}$) and central value $I_c$ of the moment of inertia (in units of
$M_{\odot} \, \text{km}^2$), as well as the resulting radius ranges from
Fig.~\ref{fig:radius_constraints} (in units of km), assuming a relative
uncertainty of $\Delta I = \pm 10\%$ and $\pm 20\%$, respectively.
The last column gives the radius range in the absence of a moment of inertia
measurement. For each assumed mass, we consider three values of $I_c$ that
approximately correspond to the soft, intermediate, and stiff EOS: $I_{\text{low}}$,
$I_{\text{int}}$, and $I_{\text{high}}$, respectively
\end{table}

\begin{figure*}[t]
\centering
\includegraphics[width=\textwidth]{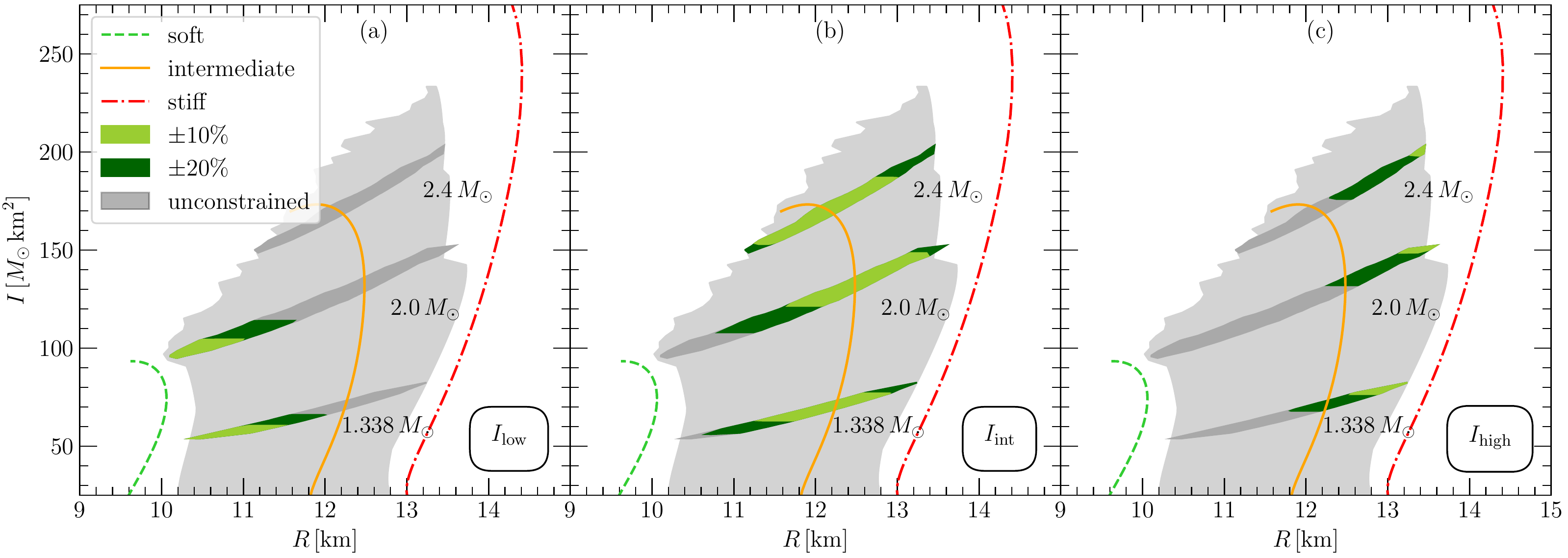}
\caption{Same as Fig.~\ref{fig:radius_constraints} but using the speed of sound
model from \citet{Greif19MNRAS} to extrapolate to high densities.}
\label{fig:radius_constraints_CS}
\end{figure*}

\begin{table}[t]
\centering
\caption{Same As Table~\ref{tab:radius_ranges} but Corresponding to 
Fig.~\ref{fig:radius_constraints_CS} Using the Speed of Sound
Model to Extrapolate to Higher Densities}
\label{tab:radius_ranges_CS}
\begin{tabular}{cccccc}
$M$ & & $I_c$ & $R(\pm 10\%)$ & $R(\pm 20\%)$ & $R$ \\ 
\hline \hline
& $I_{\text{low}}$ & 55 & 10.4--11.5 & 10.4--12.0 & 10.4--13.2 \\
1.338 & $I_{\text{int}}$ & 70 & 11.3--12.9 & 10.7--13.2 & 10.4--13.2 \\ 
& $I_{\text{high}}$ & 85 & 12.6--13.2 & 11.8--13.2 & 10.4--13.2 \\ \hline
& $I_{\text{low}}$ & 95 & 10.1--11.0 & 10.1--11.6 & 10.1--13.6 \\ 
2.0 & $I_{\text{int}}$ & 135 & 11.7--13.4 & 10.8--13.6 & 10.1--13.6 \\ 
& $I_{\text{high}}$& 165 & 13.2--13.6 & 12.3--13.6 & 10.1--13.6 \\ \hline
& $I_{\text{low}}$ & -- & -- & -- & 11.1--13.5 \\ 
2.4 & $I_{\text{int}}$ & 170 & 11.3--13.0 & 11.1--13.5 & 11.1--13.5 \\ 
& $I_{\text{high}}$ & 220 & 13.3--13.5 & 12.3--13.5 & 11.1--13.5 \\ \hline \hline
\end{tabular} 
\end{table}

\section{Improved constraints from moment of inertia measurements}
\label{sect:constraints}

Based on the frameworks discussed in Section~\ref{sect:MOI_EFT}, we now
investigate to what extent moment of inertia measurements
can improve these constraints. To this end, we assume that it is possible to
measure simultaneously the neutron star mass (with negligible uncertainty)
and the moment of inertia with central value $I_c$ and relative uncertainty
of $\Delta I = \pm 10 \%$ and $\pm 20 \%$, respectively. We
consider three different masses, $M=1.338 \, M_{\odot}$, $2.0 \, M_{\odot}$,
and $2.4 \, M_{\odot}$, and for each mass, three possible central values
$I_c$, given by $I_{\text{low}}$,
$I_{\text{int}}$, and $I_{\text{high}}$, which approximately correspond to the moment of inertia given by
the three representative EOSs shown in panel~(c) of Fig.~\ref{fig:MRI}. The
values of $I_c$ for these assumed measurements are listed in 
Table~\ref{tab:radius_ranges}, where we also give the improved radius
ranges resulting from such a simultaneous measurement. In addition,
we show the allowed $I$--$R$ areas in Fig.~\ref{fig:radius_constraints},
where the three panels correspond to the low, intermediate, and high
$I_c$ cases. For a $2.4 \, M_\odot$ neutron star, the soft EOS
is ruled out (see Fig.~\ref{fig:MRI}), and no low $I_c$ scenario exists
in this case. We also note that the EOS can have a more intricate
behavior in the general EOS band, e.g, going from soft to stiff and vice
versa with higher slopes in the $M$--$R$ diagram (see Fig.~\ref{fig:MRI_extremes}).

Moreover, we show in Table~\ref{tab:radius_ranges_CS} and 
Fig.~\ref{fig:radius_constraints_CS} how these radius constraints change
if one uses the speed of sound model instead of the piecewise polytrope extension. The results show that the radius 
constraints are remarkably consistent, with the largest differences
due to the underlying allowed bands (see the gray regions versus
the area within the representative EOS in Fig.~\ref{fig:radius_constraints_CS}),
occurring for heavy mass neutron stars and
the high $I_c$ case.

\begin{figure*}[t]
\centering
\includegraphics[width=0.7\textwidth]{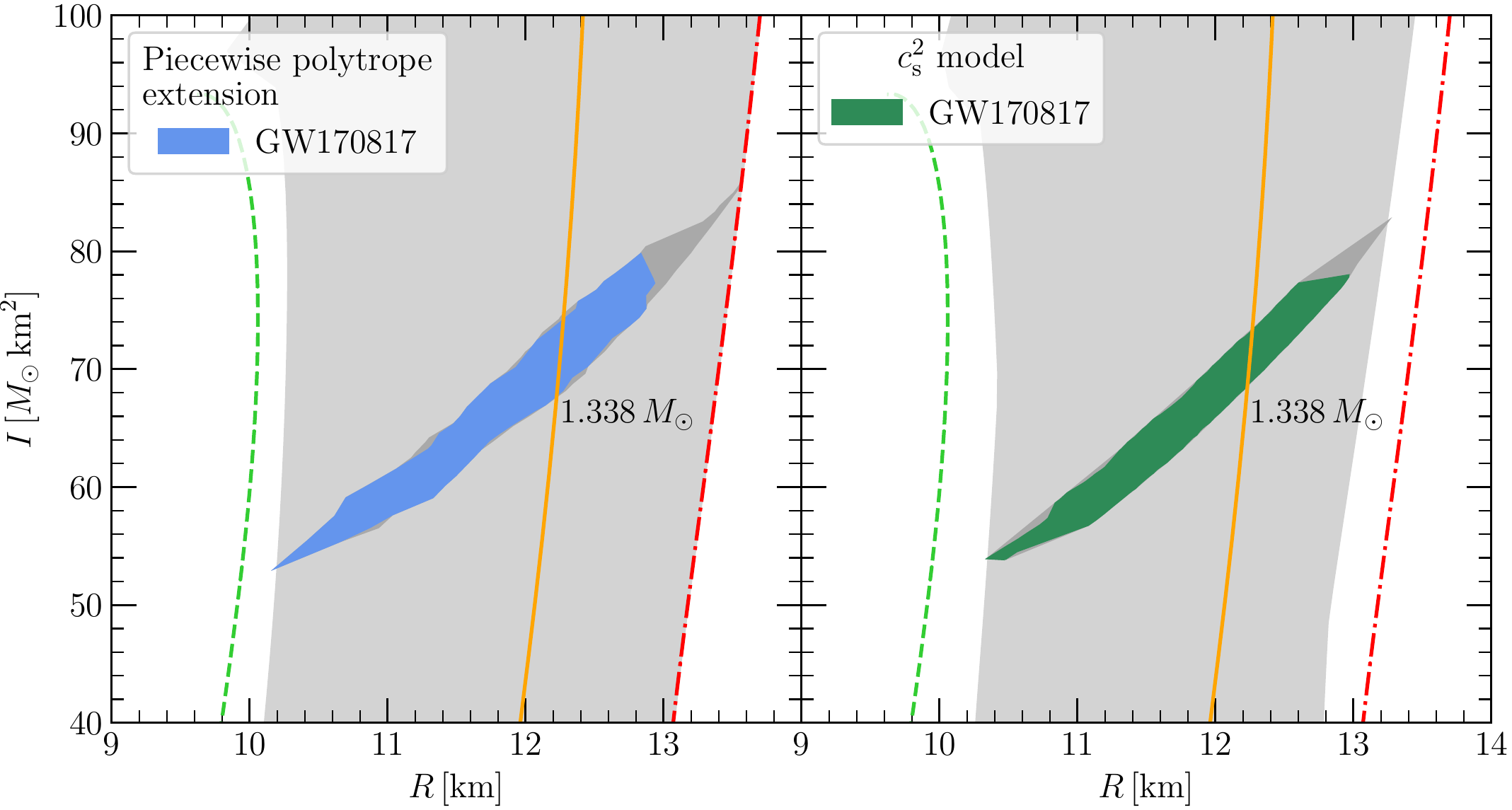}
\caption{Allowed values for the moment of inertia $I$ as a function of radius $R$ (gray
bands) resulting from the general EOS construction using the piecewise polytrope 
extension (left panel) and the speed of sound model (right panel). The darker gray 
regions indicate the  $I-R$ pairs that are consistent with a $1.338 \, M_{\odot}$ 
neutron star, whereas the blue and green highlighted areas include in addition
the GW170817 constraints for the chirp mass, mass ratio, and binary tidal deformability
from LIGO/Virgo \citep{LIGO19update}, see the text for details.}
\label{fig:radius_M1338_GW170817}
\end{figure*}

Figures~\ref{fig:radius_constraints} and~\ref{fig:radius_constraints_CS} 
clearly show that a measurement of
$I_c$ with a relative uncertainty of $\Delta I = \pm 10 \%$ ($\pm 20 \%$)
in (almost) all cases significantly improves the constraints on neutron star
radii. For a $\pm 10 \%$ measurement, if the measured value of $I_c$ is
located close to the center of the EOS band, the radius range decreases
by about 50\%, whereas the radius becomes even more narrowly
predicted when $I_c$ is close to low or high values. In the latter cases, 
the radius spread in Table~\ref{tab:radius_ranges} is only 0.9--1.2\,km for 
the piecewise polytrope extension and 0.2--1.1\,km for the speed of sound model.

\begin{figure*}[t]
\centering
\includegraphics[width=\textwidth]{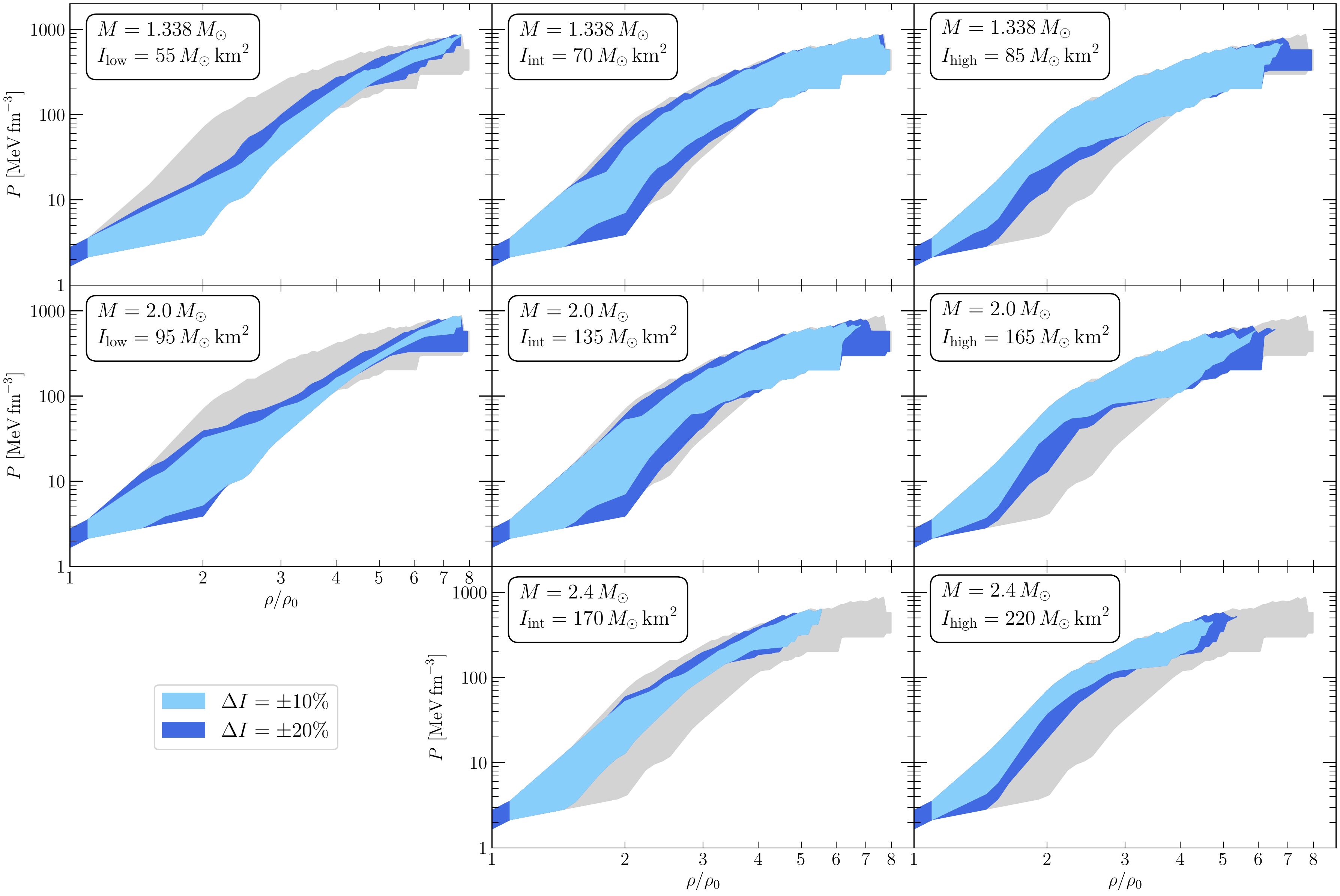}
\caption{Pressure $P$ as a function of mass density $\rho/\rho_0$ in units
of the saturation density. The gray region is the general EOS band based on
the piecewise polytrope extension. The light and dark blue
areas show the allowed EOS range for assumed simultaneous measurements 
of the mass (different rows) and the moment of inertia (different columns), as in 
Fig.~\ref{fig:radius_constraints} and Table~\ref{tab:radius_ranges},
with a relative uncertainty of $\Delta I = \pm 10\%$ $(\pm 20\%)$.}
\label{fig:EOS_constraints}
\end{figure*}

\begin{figure*}[t]
\centering
\includegraphics[width=1.\textwidth]{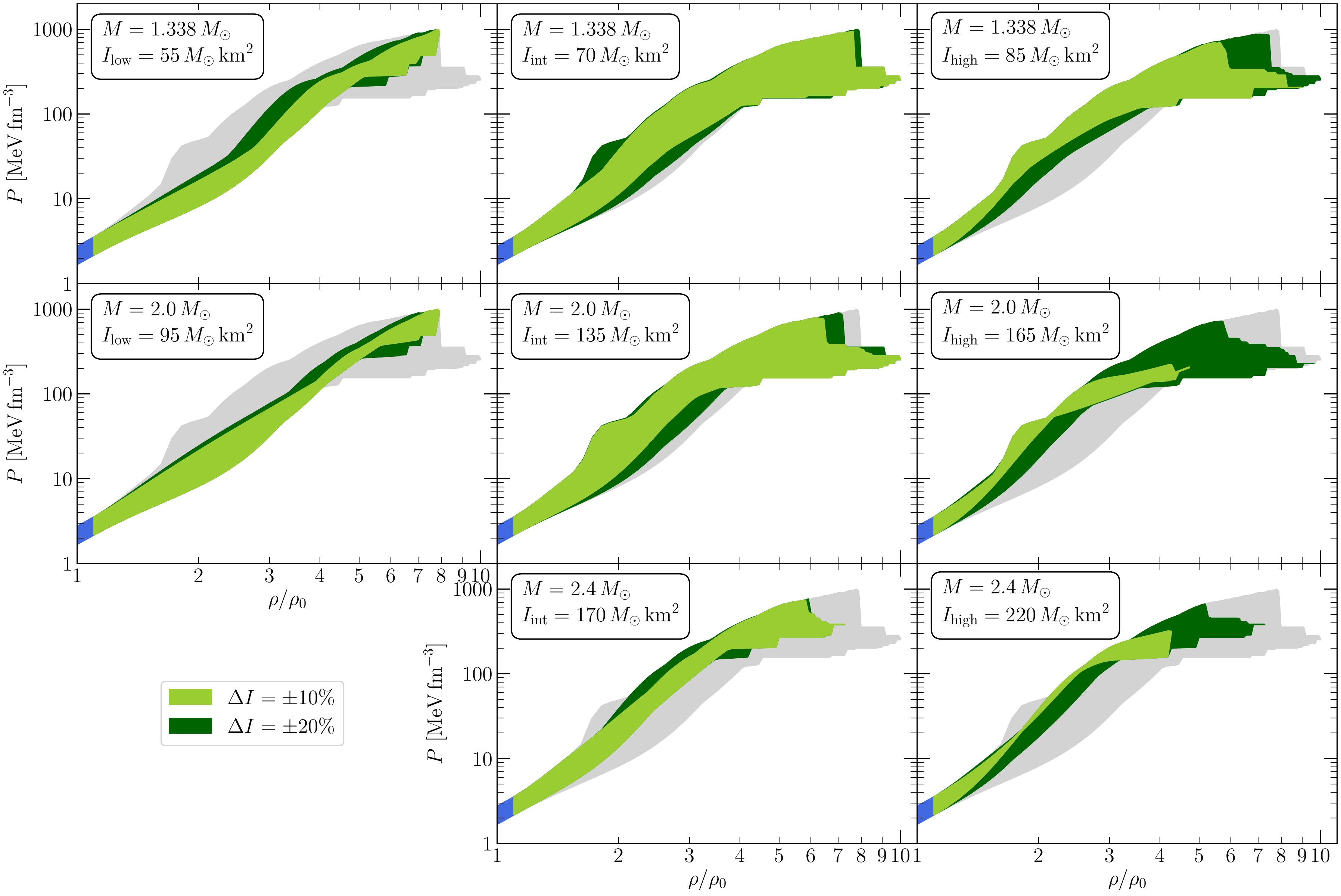}
\caption{Same as Fig.~\ref{fig:EOS_constraints} but using the speed of sound
model, corresponding to Fig.~\ref{fig:radius_constraints_CS} and 
Table~\ref{tab:radius_ranges_CS}.}
\label{fig:EOS_constraints_CS}
\end{figure*}

Next, we focus on the neutron star PSR J$0737-3039$A with mass $1.338 \, M_{\odot}$, 
which is the target of a future moment of inertia measurement. In 
Fig.~\ref{fig:radius_M1338_GW170817} we show the allowed values for the
moment of inertia as a function of radius resulting from the piecewise polytrope 
extension (left panel) and the speed of sound model (right panel), where the darker
gray regions indicate the  $I-R$ pairs that are consistent with a $1.338 \, M_{\odot}$ 
star. The impact of an accurate $I$ measurement is clear from the representative
cases in Tables~\ref{tab:radius_ranges} and~\ref{tab:radius_ranges_CS}. 
Figure~\ref{fig:radius_M1338_GW170817} shows again that the tightest
radius constraints would result from $I_c$ values toward the extremes of our 
general EOS bands.

\begin{figure}[t]
\centering
\includegraphics[width=0.45\textwidth]{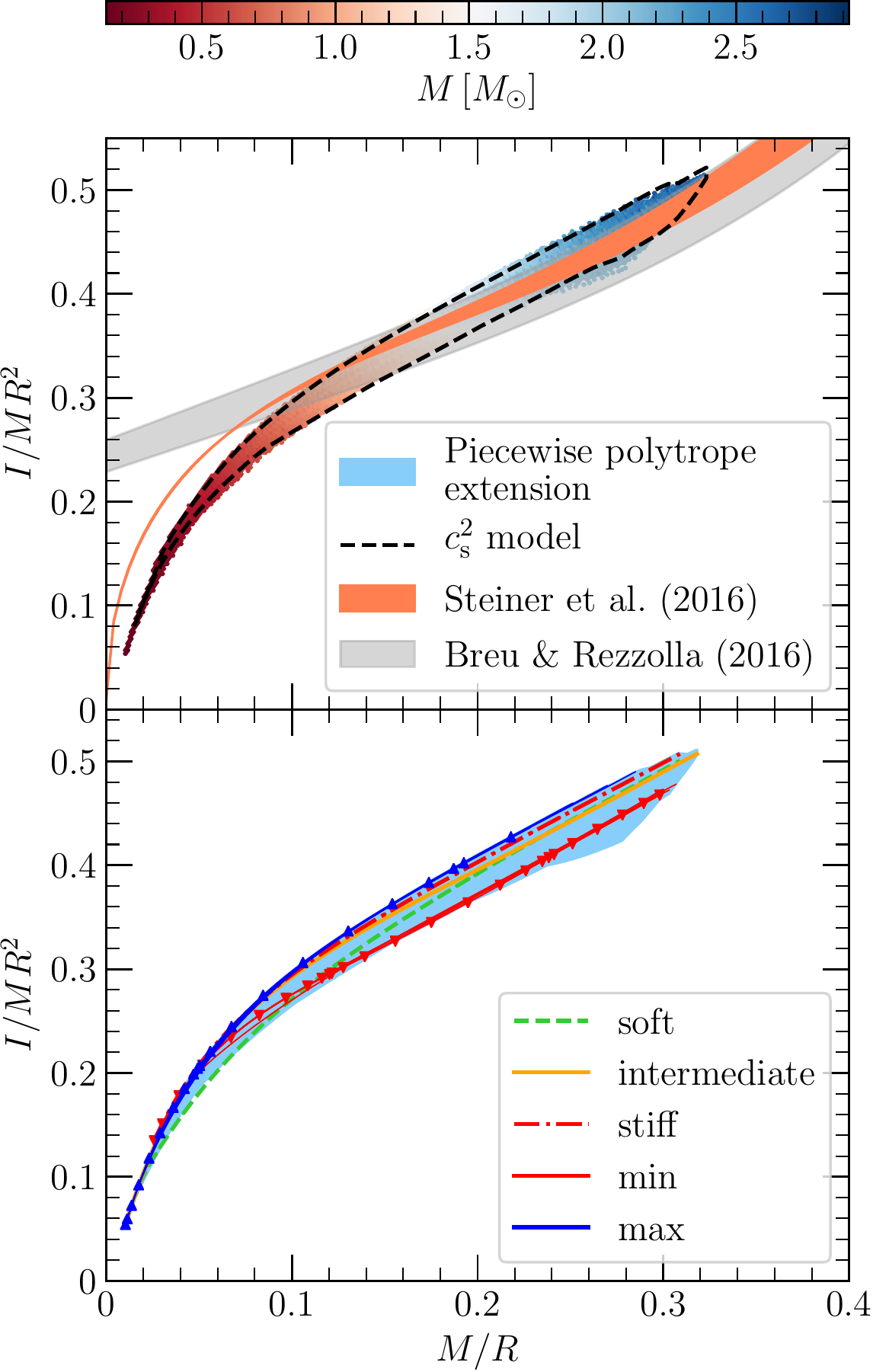}
\caption{Dimensionless moment of inertia $I/MR^2$ as a function of
compactness $M/R$. The red-to-blue region in the upper panel and the
light blue region in the lower panel show our results for the general EOS
band using the piecewise polytrope extension, with color coding according
to the neutron star mass in the upper panel. In addition, we also show
the results for the speed of sound model as the region enclosed by the
black dashed lines. In the upper panel, this is compared to correlation bands from 
\citet{Stei16unirel} in orange as well as \citet{Breu16nsMOI} in gray. 
In the lower panel, we also show the three representative EOS (soft,
intermediate, and stiff) of \citet{Hebe13ApJ}. The red (blue) lines with down
(up) triangle points are the individual EOS within the piecewise polytrope
extension with minimal $\chi^2$ of $I/MR^2$ with respect 
to the lower (upper) boundary (from fits for $M/R \geqslant 0.1$).}
\label{fig:dimless_I}
\end{figure}

\begin{figure}[t]
\centering
\includegraphics[width=0.45\textwidth]{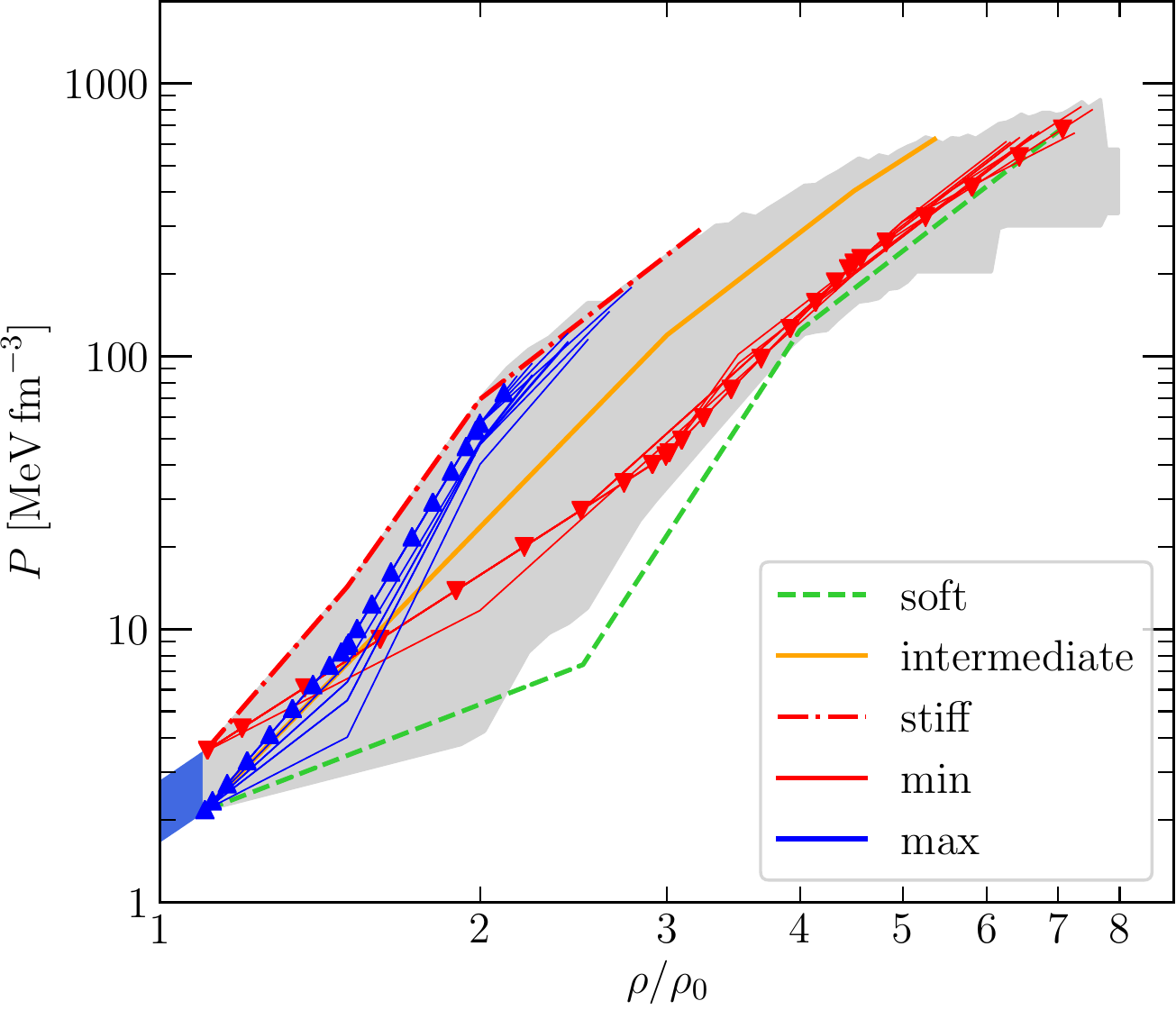}
\caption{Pressure $P$ as a function of mass density $\rho/\rho_0$ in units
of the saturation density. The gray region is the general EOS band based on
the piecewise polytrope extension. The lines correspond
to the individual EOS shown in the lower panel of Fig.~\ref{fig:dimless_I}, 
where the red and blue lines extremize the $I/MR^2$--compactness
correlation.}
\label{fig:extreme_I}
\end{figure}

In addition, we explore the constraints from the gravitational-wave signal of the
neutron star merger GW170817~\citep{LIGO18NSradii, LIGO19update}. In 
Fig.~\ref{fig:radius_M1338_GW170817}, we have highlighted the $I-R$ regions
in blue (green) for the general EOS construction based on the piecewise polytrope
extension (speed of sound model) that are consistent with the LIGO/Virgo 
results~\citep{LIGO19update} for
the  chirp mass $\mathcal{M} = 1.186 \pm 0.001 \, M_{\odot}$, 
the mass ratio $q = 0.73 - 1.00$, and 
the binary tidal deformability $\widetilde{\Lambda} = 300^{+ 420}_{- 230}$
(for the 90\% highest posterior density interval). These ranges are compatible with the 
analysis of \citet{De18tidaldef}, suggesting that they are robust with respect to assumptions 
about the underlying EOS and deformability priors. The comparison to the general EOS 
regions without the GW10817 constraints (darker gray vs.~blue and green regions) in 
Fig.~\ref{fig:radius_M1338_GW170817} shows that the GW170817 observation is 
consistent with the general EOS band based on nuclear physics and the observation of 
$2 \, M_\odot$ neutron stars. 

\begin{figure*}[t]
\centering
\includegraphics[width=\textwidth]{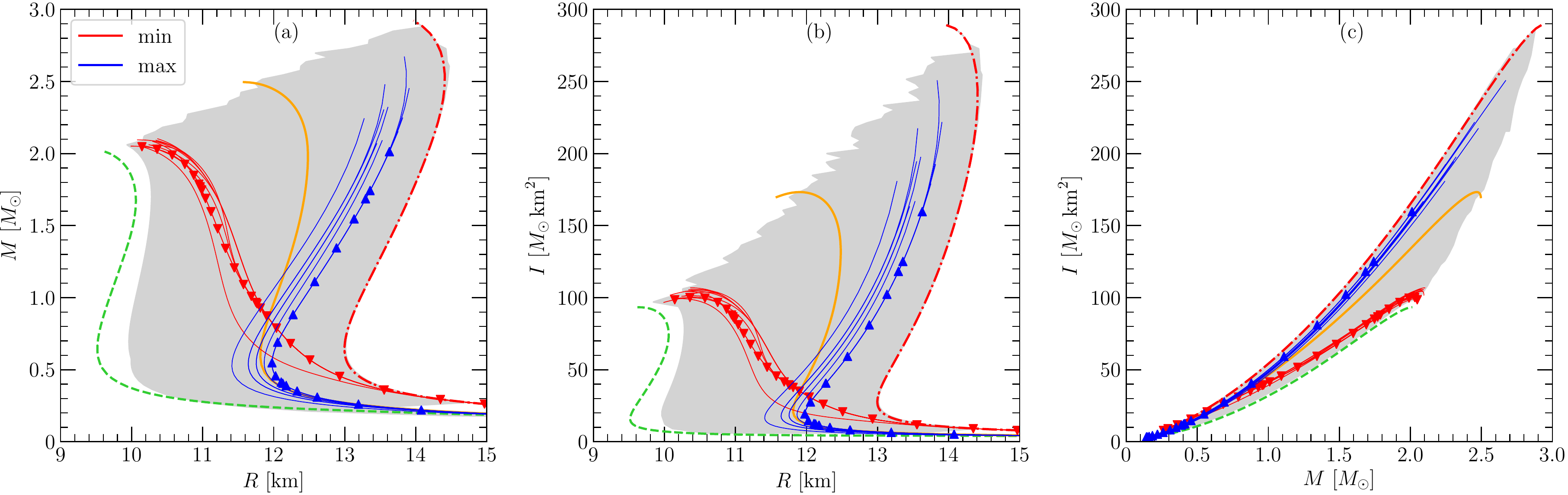}
\caption{Same as Fig.~\ref{fig:MRI}, but including the individual EOS 
shown in the lower panel of Fig.~\ref{fig:dimless_I}, where the red and 
blue lines extremize the $I/MR^2$--compactness correlation.}
\label{fig:MRI_extremes}
\end{figure*}

In addition to the radius constraints based on a moment of inertia measurement,
we can also study the corresponding constraints for the EOS. The different $I_c$
and mass scenarios for the piecewise polytrope extension (corresponding to the
radius constraints of Fig.~\ref{fig:radius_constraints} and Table~\ref{tab:radius_ranges}) are shown in 
Fig.~\ref{fig:EOS_constraints}. The gray region is again the general EOS band 
of \citet{Hebe13ApJ} (updated for the maximum mass constraint), whereas the 
different panels show the constraints for the assumed simultaneous measurements 
of the mass (different rows) and the moment of inertia (different columns).
Naturally, we find that the constraints on the EOS are the strongest for those cases
that also give the strongest radius constraints. In addition, small values 
of $I$ tend to give stronger constraints on the EOS at higher densities, 
whereas large values for $I$ provide stronger constraints at lower densities.
Moreover, measurements of heavy neutron stars provide stronger constraints
on the EOS than the scenarios for typical neutron stars. Further, we give
in Fig.~\ref{fig:EOS_constraints_CS} the EOS constraints for the speed of
sound model (corresponding to the radius constraints of 
Fig.~\ref{fig:radius_constraints_CS} and Table~\ref{tab:radius_ranges_CS}).
This shows very similar constraints on the EOS, as for the piecewise
polytrope extension.

Several studies based on different phenomenological EOS have shown
that the dimensionless moment of inertia $I/M R^2$ correlates with the 
compactness $M/R$ to a good approximation 
\citep{Latt00ns,Bejg02Crab, Latt05nsMOI, Breu16nsMOI}. In
Fig.~\ref{fig:dimless_I} we present our results for the piecewise polytrope
extension (color coded) and the speed of sound model (black dashed line)
for the dimensionless moment of inertia, which yield a very similar correlation
band, and compare these to the bands from \citet{Stei16unirel} and 
\citet{Breu16nsMOI}. Our results agree reasonably well with these for 
$M/R > 0.15 \, M_\odot/$km, while we find a deviation for smaller 
compactness parameters and also a somewhat larger band for
$M/R > 0.2 \, M_\odot/$km. This shows that, e.g., predictions for neutron 
stars with small mass and large radii based on the former correlation
bands are not compatible with the general EOS band. This is most
likely due to low-density assumptions made that are incompatible 
with modern nuclear physics.

In addition, we show in the lower panel of Fig.~\ref{fig:dimless_I}
the three representative EOSs (soft, intermediate, and stiff) of \citet{Hebe13ApJ}.
These are representative with respect to radius and moment of inertia 
for all masses (see Fig.~\ref{fig:MRI}) but, as is clear from Fig.~\ref{fig:dimless_I}, they
do not capture the extremes of the dimensionless moment of inertia.
In order to investigate the band for the dimensionless moment of inertia
in more detail, we determined the individual EOSs that represent the
limits of the band in Fig.~\ref{fig:dimless_I} for the piecewise polytrope
extension, which provides the more conservative estimate. To this end, 
we discretized $M/R$ for $M/R \geqslant 0.1 \, M_\odot\,$km$^{-1}$ and determined
the $\chi^2$ of each EOS for the deviation of $I/M R^2$ from the lower
(upper) band. The results for the individual EOSs with the minimal 
$\chi^2$ values are shown as red (blue) lines in the lower panel of 
Fig.~\ref{fig:dimless_I}.

The corresponding EOSs for these extreme cases are shown in
Fig.~\ref{fig:extreme_I}. We observe that the EOSs with a minimum 
$\chi^2$ with respect to the lower boundary of the dimensionless
moment of inertia $I/M R^2$ (red lines) 
tend to be rather stiff at nuclear densities and soft at high densities, 
whereas the EOSs leading to large values of $I/M R^2$ tend to be soft 
at nuclear densities and stiff at high densities (blue lines). These 
trends are also reflected in the results for the mass, radius, and 
moment of inertia in Fig.~\ref{fig:MRI_extremes}, where these individual
EOSs are clearly extreme but nevertheless very interesting cases. 
The EOSs with the low
values for the dimensionless moment of inertia predict large radii 
at small masses (and moment of inertia) and small radii at larger 
masses (red lines), while the ones corresponding to large values for
$I/M R^2$ show the opposite trend.

\section{Summary and Outlook}
\label{sect:summary}

We have explored new and improved constraints for the EOS of neutron-rich
matter and neutron star radii. Our work is based on four inputs: (a) 
microscopic calculations of the EOS up to $1.1 \, \rho_0$ 
based on state-of-the-art nuclear interactions derived from chiral EFT
combined with the piecewise polytrope or speed of sound extension
to high densities following \citet{Hebe13ApJ} and \citet{Greif19MNRAS},
respectively, (b) the precise measurement of the mass of PSR J0740+6620 with
$2.14 \substack{+0.10 \\ -0.09} \, M_\odot$~\citep{Crom19massive},
(c) causality constraints at all densities and an asymptotic behavior of
the speed of sound consistent with perturbative QCD calculations at very
high densities for the $c_\text{s}^2$ model, and (d) constraints from future 
measurements of the mass and 
moment of inertia of the same star. Note that this analysis does not rely 
on any assumptions regarding the composition and properties of 
matter beyond the density $1.1 \, \rho_0$, and within the space
of the piecewise polytrope and speed of sound extension includes
EOS that mimic regions with a first-order phase transition. 

For the moment of inertia measurements we considered different 
scenarios by assuming various values and uncertainties for the 
moment of inertia. We find that measurements with an uncertainty 
of $10 \%$ lead to a reduction of the radius range by about 50\%
compared to the general EOS band from \citet{Hebe13ApJ} and \citet{Greif19MNRAS}
when the moment of inertia corresponds to an intermediate EOS.
If the moment of inertia corresponds to values predicted by a soft 
or stiff EOS the radius range is reduced by a factor of 3 or more.
For all $\pm 10\%$ measurements, the resulting radius range is 
smaller than $1.9\,$km for all considered masses $M=1.338, \ 2.0$, 
and $2.4 \, M_{\odot}$. Specifically, for a $1.338 \, M_{\odot}$ star,
we find radius ranges of $R=10.2$--$11.5\,$km for low values of the
moment of intertia ($I_\text{low}=55 \, M_{\odot}\,$km$^2$ with $\Delta I = \pm 10 \%$;
combining the ranges from the piecewise polytropic and speed of sound extensions),
$R=11.3$--$12.9\,$km for intermediate values ($I_\text{int}=70 \, M_{\odot}\,$km$^2$),
and $R=12.5$--$13.6\,$km for high values ($I_\text{high}=85 \, M_{\odot}\,$km$^2$).
These ranges need to be compared with $R=10.2$--$13.6\,$km based
on the combined general EOS bands for this mass, when no information
about the moment of inertia is used. We have also
investigated the corresponding constraints for the EOS. We found that
large values for the moment of inertia provide stronger constraints at lower
densities, whereas small values tend to constrain the EOS at higher densities.
Moreover, measurements of heavy neutron stars provide overall stronger
constraints. In addition, we have studied the dimensionless
moment of inertia $I/M R^2$ and established the full uncertainty
ranges based on our general piecewise polytrope and speed of
sound extension. We find very interesting extreme EOSs at the 
boundaries of the correlation with the compactness, which have
not been considered before.

Finally, we showed that the gravitational-wave constraints from the
neutron star merger GW170817~\citep{LIGO18NSradii, LIGO19update}
are consistent with the general EOS bands explored here (see also
\citet{Raai20NICERGW}). We found that the latest analysis of GW170817 
\citep{LIGO19update} only slightly reduces the radius range predicted
by the general EOS bands from the piecewise polytrope and speed of
sound extension, and only weakly narrows the range for the predicted 
moment of inertia for a $1.338 \, M_{\odot}$ star. Therefore,
additional future detections from LIGO/Virgo, as well as NICER and 
other X-ray timing observations~\citep{Watt16RMP}, combined with 
measurements of neutron star masses and in particular the moment 
of inertia, are a powerful avenue to further constrain the EOS 
of dense matter in a model-independent way.

\section*{Acknowledgments}

We thank G.\ Raaijmakers, I.\ Tews, and A.\ Watts for useful discussions. This 
work was supported in part by the Deutsche Forschungsgemeinschaft (DFG, 
German Research Foundation) -- Project-ID 279384907 -- SFB 1245, and 
J.M.L. acknowledges support from NASA through grant 80NSSC17K0554 
and the U.S. DOE from grant DE-FG02-87ER40317.

\bibliographystyle{aasjournal}
\bibliography{strongint.bib} 

\end{document}